\documentclass[pdflatex,sn-mathphys-num]{sn-jnl}% Math and Physical Sciences Numbered Reference Style

\usepackage{graphicx}%
\usepackage{multirow}%
\usepackage{amsmath,amssymb,amsfonts}%
\usepackage{amsthm}%
\usepackage{mathrsfs}%
\usepackage[title]{appendix}%
\usepackage[table]{xcolor}%
\usepackage{textcomp}%
\usepackage{manyfoot}%
\usepackage{booktabs}%
\usepackage{algorithm}%
\usepackage{algorithmicx}%
\usepackage{algpseudocode}%
\usepackage{listings}%
\usepackage{tabularx,booktabs,multirow, array} % comparison to openebench, openproblems
\usepackage{lineno}
\usepackage[many]{tcolorbox}
\tcolorboxenvironment{lstlisting}{
  spartan,
  breakable,
  frame empty,
  boxsep=0mm,
  boxrule=0pt,
  left=1mm, right=1mm, top=-1mm, bottom=-1mm,
  colback=gray!15,
}

\lstalias{yaml}{}   % allow language=yaml
\usepackage{geometry}

\usepackage{setspace}
\begin{document}

\title[Omnibenchmark]{Omnibenchmark: transparent, reproducible, extensible and standardized orchestration of solo and collaborative benchmarks}

\author*[1,2]{\fnm{Izaskun} \sur{Mallona}}\email{izaskun.mallona@gmail.com}

\author[3]{\fnm{Almut} \sur{Lütge}}
\author[1,2]{\fnm{Ben} \sur{Carrillo}}
%\email{iiauthor@gmail.com}

\author[1]{\fnm{Daniel} \sur{Incicau}}
%\email{iiiauthor@gmail.com}

\author[1,2]{\fnm{Reto} \sur{Gerber}}
\author[1]{\fnm{Aidan} \sur{Meara}}
\author[1,2]{\fnm{Anthony} \sur{Sonrel}}
\author[2,4]{\fnm{Charlotte} \sur{Soneson}}
\author*[1,2]{\fnm{Mark D.} \sur{Robinson}}\email{mark.robinson@mls.uzh.ch}

\affil[1]{Department of Molecular Life Sciences, University of Zurich, Zurich, Switzerland}
\affil[2]{SIB Swiss Institute of Bioinformatics, Zurich, Switzerland}
\affil[3]{Swiss Data Science Center, Zurich, Switzerland}
\affil[4]{Friedrich Miescher Institute for Biomedical Research, Basel, Switzerland}

\abstract{Benchmarking involves designing, running and disseminating rigorous performance assessments of methods, most often for data analysis and software tools, but the process can also be applied to experimental systems. Ideally, a benchmarking {\em system} is used to facilitate the benchmarking process by providing a structured entrypoint to design, coordinate, execute, and store standardized benchmarks. We describe a novel benchmarking system, \texttt{Omnibenchmark}, that facilitates benchmark formalization and execution in both solo and community efforts.  \texttt{Omnibenchmark} provides a flexible benchmark plan syntax (i.e., a configuration \texttt{YAML} file), dynamic workflow generation based on \texttt{Snakemake}, S3-compatible storage handling, and reproducible software environments using environment modules, \texttt{Apptainer} or \texttt{Conda}. Such a setup provides an unprecedented flexibility such that existing benchmark designs can be forked and extended, run separately or collaboratively, giving versioned and standardized result outputs and therefore much-needed transparency to the analysis and interpretation of benchmark results. Tutorials and installation instructions are available from \texttt{https://omnibenchmark.org}.}


\maketitle

Computational tools are a staple in all endeavours of scientific research involving data. Researchers use a variety of computational tools to process and interpret their data and/or to refine these methods and develop new ones; often there are competing tools for each step of an analysis pipeline. Scientific benchmarking provides a means to compare method performances, to pinpoint strengths and weaknesses and to help choose well-suited methods for a given purpose and context.

Benchmarking strategies and subfield traditions vary widely. Benchmarks are typically carried out by individuals or small teams, can be neutral or not, and are often not updated after publication; a particularly large class of such benchmarks include those within studies that introduce new methods. Since this style of self-assessment benchmark is rarely extensible, and generally only open to scrutiny at review time, they would benefit from the use of a benchmarking system that enforces standardization and transparency. In parallel, independent teams often publish overlapping benchmarks on the same topic; however, benchmark results may not be concordant and tend to be hard to reuse due to lack of clear licensing, adoption of workflow frameworks, and limited capture of software versions and parameters \cite{cao,Sonrel2023}.

Collaborative benchmarking is an alternative to isolated (solo or small team) benchmarking \cite{openebench,seppey2020lemmi,hanssen2023ncbench,openproblems,kryshtafovych2023critical}, often using workflow systems, building communities via challenges or hackathons, sharing data using standards and FAIR \cite{wilkinson2016fair} principles, using free and open source software, tracking contributions and their neutrality, and providing infrastructure to facilitate computing in comparable and homogeneous systems (e.g., containerization). Initiatives such as the DREAM challenges \cite{stolovitzky2007dialogue} and the Critical Assessment of Structure Prediction (CASP) \cite{kryshtafovych2023critical} and of Metagenome Interpretation (CAMI) \cite{meyer2022critical} have successfully established standards and provided a neutral ground for methods comparison. The user base of community benchmarks is broad and extends beyond method developers: in bioinformatics, \texttt{OpenEBench} \cite{openebench} categorizes stakeholders into developers (typically non-neutral contributors), benchmarkers (potentially neutral contributors), end-users (anyone browsing results) and funders; each group represents different aims, motivations, and preferred access points.  In particular, community benchmarking initiatives such as \texttt{OpenProblems} \cite{openproblems} or \texttt{OpenEBench} \cite{openebench} attempt to address these barriers by channeling contributors to structured and topic-driven benchmarking events, as well as by providing computing infrastructure to standardize and run benchmarks and share their results. Beyond these general-purpose systems, many task-specific benchmarking frameworks exist, including scIB \cite{scib} (single-cell integration), CellBench \cite{cellbench} (single-cell RNA-seq methods), DANCE \cite{dance} (deep-learning single-cell and spatial tasks), and dynverse \cite{dynverse} (trajectory inference; Table~\ref{tab:topic-specific}). Unlike \texttt{Omnibenchmark}, these are targeted to specific tasks rather than orchestrating arbitrary self-deployed benchmarks.

Here, we present \texttt{Omnibenchmark}, a novel distributed and open benchmarking system to self-deploy benchmarks by using a benchmark (\texttt{YAML}) plan and its automated translation into a reproducible workflow. The tool automates and standardizes routine aspects of benchmarking (workflow generation, software environment management, versioning, provenance and authorship tracking, sharing, etc.) for both solo and collaborative benchmarks. It adds standardization and transparency to benchmarking, which are the foundation of trust in method evaluation \cite{mallona2024building}.  In particular, the \texttt{YAML} plan is meant to be transportable across computational systems, facilitating an era of open interactions with benchmarking artifacts (datasets, code, and results). \texttt{Omnibenchmark} aligns with other general-purpose orchestration continuous benchmarking systems in aim, but exhibits various design differences (Table~\ref{tab:comparison}). The \texttt{Omnibenchmark} system is intended to operate in a distributed manner, while also providing means for centralization, which has implications for operations and infrastructure (see Extended Data Section~\ref{sec:centralization} for further discussion).

\begin{figure}[hbt!]  
   \centering
      \includegraphics[width=\linewidth]{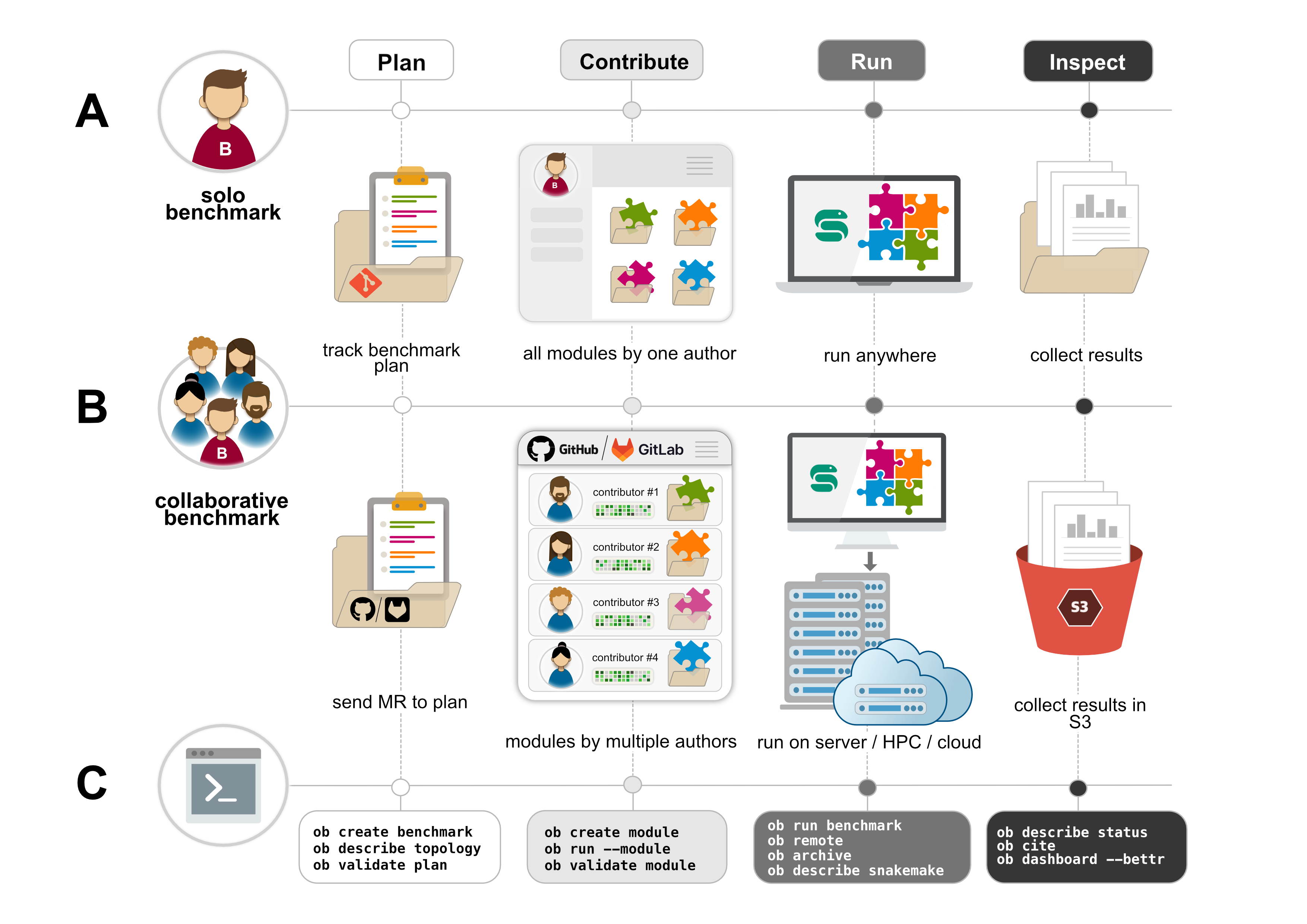} 
  \caption{The life cycle of an \texttt{Omnibenchmark}. The four main stages are depicted linearly for simplicity. \textbf{A}, the same person authors the plan and modules; results are typically kept locally. {\textbf{B}, a group collaborates on a plan via merge requests, publishes different modules, and schedules runs on reference hardware. Results are best published to remote storage. Note that the workflow can switch between solo and collaborative at any step. \textbf{C}, commands useful at each stage (full list in Table~\ref{tab:commands}). }}
  \label{fig:design}
\end{figure}

In solo benchmarking, e.g., by a person interested in comparing the performance of their method against other tools using public datasets, the researcher acts as developer, benchmarker, and end-user in their self-assessment (Figure~\ref{fig:design}A). A solo benchmark can be run locally with outputs eventually pushed to a FAIR \cite{wilkinson2016fair} repository (e.g., Zenodo) at publication time, without community building, but still facilitating transparency, extensibility, and reproducibility. Beyond solo benchmarks, \texttt{Omnibenchmark} can be used for collaborative benchmarking via hackathons, challenges, or any other community event (Figure~\ref{fig:design}B). As with solo benchmarking, the currency for community contributions is \textit{code} rather than predictions (or other locally-generated results); code is meant to be orchestrated by \texttt{Omnibenchmark} within a commonly-agreed (centralized) compute infrastructure. The operations to gather community contributions in a semi-decentralized manner resemble that of \texttt{Git} for version control: developers and benchmarkers have a local copy of the benchmark plan and required code, data, and software stacks, and can track and submit their contributions via merge requests to a commonly agreed repository on an easily accessible DevOps platform (e.g., GitHub, GitLab), from which code is collected to run the benchmark. Appropriate HPC infrastructure (e.g., virtual or physical Unix-based machines) and benchmark results (e.g., S3-compatible buckets) are defined and agreed upon by the collaboration partners, such that benchmark runs are versioned and outputs are openly accessible (ideally, also to the public). At present, the \texttt{Omnibenchmark} project does not provide compute infrastructure. Taken together, this setup makes \texttt{Omnibenchmark} flexible and novel in terms of technical features and governance (Table~\ref{tab:comparison}).

\texttt{Omnibenchmark} is a \texttt{pip}-installable Python package with functionality divided into four pillars (see Extended Data~\ref{sec:sup_methods}):

\begin{enumerate}
    \item Formalization. A \texttt{YAML} file (benchmark plan) defines all the benchmark components (Figure~\ref{fig:design}, ``Plan''; Extended Data \ref{sec:yaml}), its architecture and some of the metadata, including software environments.
      
    \item Software. The software backends are reproducibly built and activated according to the benchmark plan.  Software is specified in separate configuration files, and installed or retrieved either as \texttt{Apptainer} (formerly \texttt{Singularity}) images, local environment modules (perhaps built on \texttt{EasyBuild} \cite{hoste2012easybuild} or using and extending EESSI \cite{droge2023eessi}) or explicit \texttt{Conda} manifests.

    \item Workflow. The benchmark \texttt{YAML} plan is parsed to generate a reproducible \texttt{Snakemake} \cite{koster2012snakemake} workflow. The expressivity of the \texttt{YAML} plan enables complex benchmarks beyond data-method-metric linear layouts, including gather-scatter \cite{koster2012snakemake} patterns and adding dataset- or method-specific parameters. We use \texttt{Snakemake} (version $\geq 8$) capabilities, including its plugin ecosystem, to embrace reproducible software environments, populate remote storage, execute jobs on queuing systems, and track provenance.

    \item Storage. Final and intermediate artifacts (files) are either stored locally (solo benchmarking or during development runs) or pushed to a versioned S3-compatible storage for sharing. Benchmark releases include code, software environments, results and metadata; every release is tagged following a semantic versioning system. Write and potentially read access to S3-compatible buckets is restricted and controlled via access tokens (Figure~\ref{fig:design}, ``Inspect'').
\end{enumerate}

During execution (Figure~\ref{fig:design}, ``Run''), \texttt{Omnibenchmark} collects computational performance metrics (peak memory usage, CPU utilization, runtime, etc.) for all modules and, if part of the benchmark modules, also computes and collects algorithmic performance metrics (e.g., F1 scores, adjusted Rand indexes, etc). Both can be browsed using \texttt{bettr} \cite{bettr}, an R/Shiny application that provides interactive plots and tabular representations, or used in custom downstream analyses. 

To write a benchmark from scratch, benchmarkers need to:

\begin{enumerate}
    \item Write a benchmark \texttt{YAML} plan reflecting the modules and topology, software environments, and other metadata (Figure~\ref{fig:design}, ``Plan''). This configuration file will typically begin as a simplified version of the final benchmark and be refined iteratively. \texttt{Omnibenchmark} provides commands to template (\texttt{ob create benchmark}) and guide through the creation of a benchmark plan. The plan can be validated for syntactic and semantic conformance (\texttt{ob validate plan}); see Figure~\ref{fig:design}, ``Plan'' and  Extended Data~\ref{sec:yaml}.

    \item Populate benchmark {\em modules} as \texttt{Git}-tracked repositories (Figure~\ref{fig:design}, ``Contribute'';  Extended Data~\ref{sec:modules}). To facilitate reuse, we recommend atomic modules, i.e., each aiming to perform a single benchmarking task (e.g., simulate data, compute a distance matrix, calculate F1 scores, etc.). Each module must provide a command line interface (e.g., via \texttt{argparse}) to define input(s) and output(s), which adhere in format to the benchmark plan. \texttt{Omnibenchmark} provides a command (\texttt{ob create module}) to scaffold new benchmarking modules compatible with a given benchmark plan, and \texttt{ob validate module} for their validation.
    
    \item Define one or more software execution environments, either using \texttt{Conda}, environment modules or containers as backend; see Table~\ref{tab:software} and Extended Data~\ref{sec:operations}). The benchmark \texttt{YAML} plan allows full flexibility to run benchmarks with a software environment per module, if desired, or have multiple modules share an environment. Within a benchmark, a single backend technology is allowed.
\end{enumerate}

Running the benchmark involves issuing commands to: a) template a plan and modules; b) validate and visualize them; c) optionally, attempt a dry run; and, finally d) execute the workflow (Figure~\ref{fig:design}C). The CLI allows to submit jobs to a queuing system, or to switch from a locally-stored benchmark to cloud storage, by adding extra arguments to the command call (a full list of current commands can be found in Table~\ref{tab:commands}).

Solo benchmarks often evolve into collaborative benchmarks by adding infrastructure to (semi-)centralize the contribution, computing and sharing steps. Similar to \texttt{Git}, \texttt{Omnibenchmark} is a distributed system by nature. Every user needs to obtain a copy of the \texttt{YAML} plan, from which all required code and software dependencies can be pulled. Updating the benchmark involves changing and \texttt{Git}-tracking the \texttt{YAML} and/or referenced components, such as modules or parameters. To share changes, users must agree on a common remote (e.g., a project on GitLab, GitHub etc) and merge request changes. Other users can in turn try, approve, or reject these updates; testing such changes can be automated by continuous integration using the (GitHub) marketplace \texttt{run-omnibenchmark} action \cite{run_omnibenchmark}. Extended Data~\ref{sec:centralization} describes some infrastructure and operations that characterize different community benchmark modalities.

To call for community contributions, we recommend creating a `benchmark billboard' \texttt{Git} repository containing a thorough \texttt{README} that specifies the aim and scope of the benchmark, an initial benchmark plan and \texttt{LICENSE}, and pushing it to a remote where the community can collaboratively modify its content. Contributors would clone or pull that repository and use it locally with their \texttt{Omnibenchmark} installation. Changes to the benchmark plan (e.g., new datasets) are then pushed (via merge requests) to this repository, so that the project members in charge of running the benchmark on appropriate hardware (local, cloud or HPC) can pull the new plan, test and approve it, and run the new benchmark version and populate the object storage with the benchmark version results; meanwhile, community members can at any time fork and further develop the benchmark. We demonstrate such an extension, adding a module and rerunning with incremental reuse, in Supplementary Note~\ref{sec:extend}.

Ensuring neutral benchmarks and constructive benchmarking communities, however, goes far beyond the technical implementation. We suggest a `transparency builds trust' approach \cite{mallona2024building} as well as a code of conduct. Similarly, the \texttt{README} should be explicit about benchmark philosophy, responsibilities and costs (e.g., who provides the storage and computing infrastructure), planned changes in benchmark layout, parameter handling strategies, and desired strategy for academic publication and authorship attribution. Such documentation aligns well to pre-registration (e.g., as done in simulation studies \cite{siepe2023simulation}).

To showcase the use of \texttt{Omnibenchmark}, we have ported parts of the \texttt{clustering-benchmarks} framework \cite{gagolewski2022framework} and the spatially variable genes (SVG) benchmark \cite{bioRxiv2023spatially_variable_genes} from \texttt{OpenProblems}. 

\begin{figure}
    \centering
    \includegraphics[width=\linewidth]{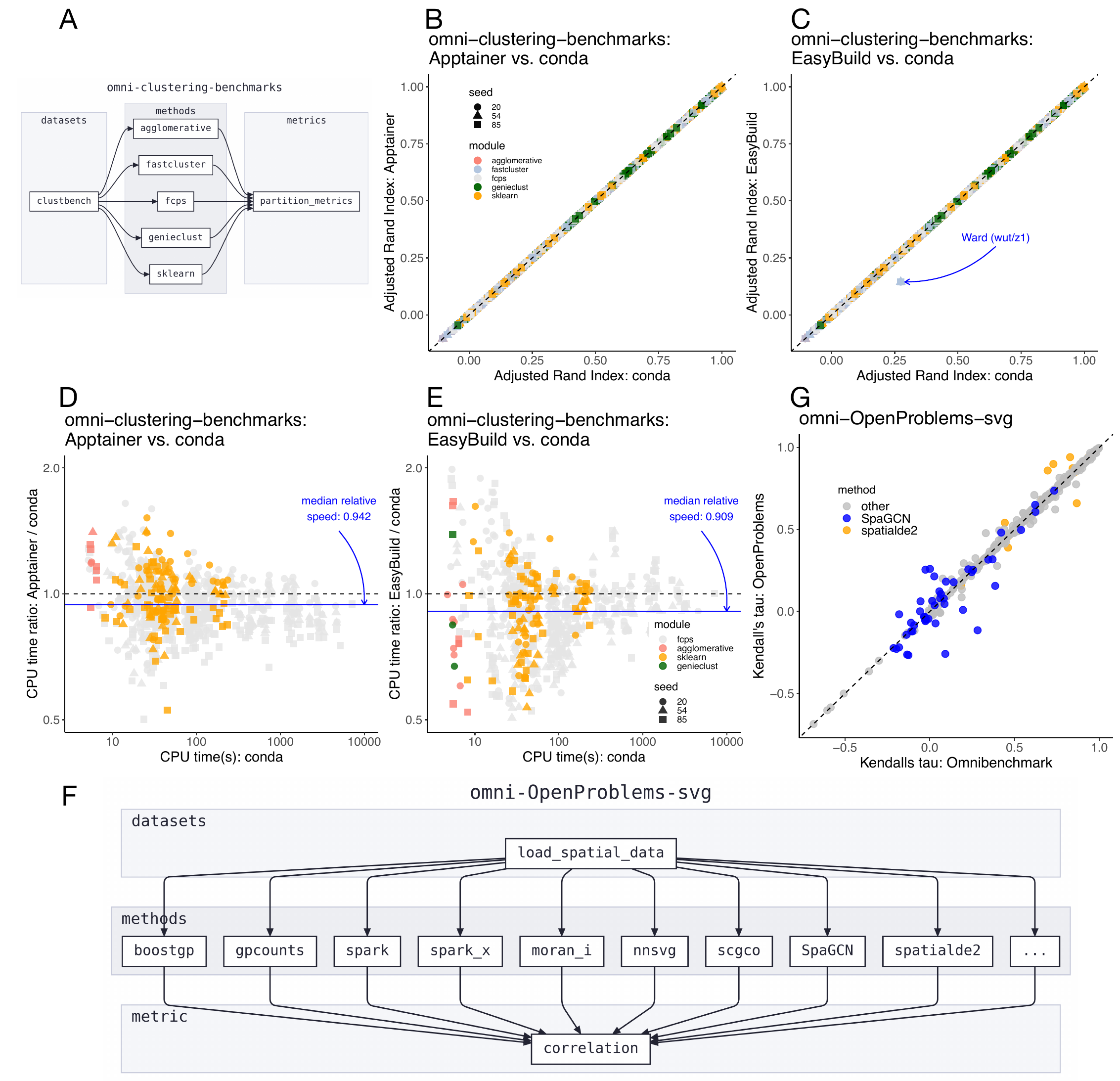}
    \caption{Implementation of the \texttt{omni-clustering-benchmarks} framework and porting the \texttt{OpenProblems} Spatially Variable Genes benchmark.  \textbf{A}, simplified benchmark plan for \texttt{omni-clustering-benchmarks}; note that each \textit{module} at each \textit{stage} implements multiple datasets, methods, or metrics via parameters (in total here, 62 datasets, 48 methods and 3 metrics are included in \texttt{omni-clustering-benchmarks}; this benchmark plan was run 18 times, for 3 seeds, 2 runs, and 3 software backends). \textbf{B}, computed clustering performance (Adjusted Rand Index) with \texttt{Apptainer} software versus \texttt{Conda} software. \textbf{C}, computed clustering performance (Adjusted Rand Index) with \texttt{EasyBuild} software versus \texttt{Conda} software. \textbf{D}, relative CPU time of \texttt{Apptainer} software compared to \texttt{Conda} software versus \texttt{Conda} CPU time. \textbf{E}, relative CPU time of \texttt{EasyBuild} software compared to \texttt{Conda} software versus \texttt{Conda} CPU time. \textbf{F}, simplified benchmark plan for \texttt{omni-OpenProblems-svg}; here, the \texttt{load\_spatial\_data} module parameterizes 16 datasets, each \texttt{methods} module implements a single method; the \texttt{correlation} module implements Kendall's $\tau$ against the ground truth. \textbf{G}, \texttt{omni-OpenProblems-svg} reproduces to a high degree the Kendall's $\tau$ values computed in the original \texttt{OpenProblems} benchmark.}
    \label{fig:example_benchmarks}
\end{figure}

The original \texttt{clustering-benchmarks} framework wraps code, data, simulation and metric efforts from a whole community to highlight the performance of clustering algorithms across a variety of datasets \cite{gagolewski2022framework}, extending the traditional UCI datasets and their expertly-curated ground truths with extra datasets \cite{franti2018k,graves2010kernel,ultsch2005clustering, gagolewski2021cluster}. We ported a large portion of the \texttt{clustering-benchmarks} framework as an Omnibenchmark, denoted here as \texttt{omni-clustering-benchmarks} (simplified plan in Figure \ref{fig:example_benchmarks}A) and ran it 18 times (3 software backends, 3 seeds, 2 runs), with a focus on three aspects: reproduction of results across seeds and runs from each software backend (i.e., \texttt{Apptainer}, \texttt{Conda}, \texttt{EasyBuild}-built environment modules); analysis of computational performance across backends; and, showing the flexibility of \texttt{Omnibenchmark} by running parameter sweeps (e.g., cluster number misspecification). We found that, even with slight differences in package version specifications, the clustering performance metrics were nearly identical across backends if propagating the seeds accordingly (Figure~\ref{fig:example_benchmarks}B-C); although top-level versions are consistent for the \texttt{agglomerative} module, one reproducible deviation exists for Ward's method applied to the \texttt{wut-z1} dataset for \texttt{EasyBuild}-built software and we isolated this discrepancy to originate from a compiler flag (see Extended Data \ref{sec:wut-z1-ward}).  Differences in CPU time were observed depending on the backend (Figure~\ref{fig:example_benchmarks}D-E), highlighting modest speed gains for \texttt{Apptainer} compared to \texttt{Conda} (median: 3.0-5.8\% faster depending on filters; see Extended Data \ref{fig:cpu_time-supp}) and more substantial gains by running tools on \texttt{EasyBuild}-built software (median: 8.7-9.1\% faster; Extended Data \ref{fig:cpu_time-supp}). In contrast, memory usage was higher for \texttt{Apptainer} (median: 12.3\%) and modestly lower for \texttt{EasyBuild} (median: 2.9\%) relative to \texttt{Conda} (Extended Data \ref{fig:max_rss-supp}). Lastly, it is clear that misspecifying the number of clusters leads to a decrease in algorithmic performance, on average (Extended Data \ref{fig:cluster-misspec-supp}-\ref{fig:cluster-misspec-supp1}).

%As for the consequences of misspecifying the (true) number of clusters within the clustering method calls, we found some methods were more robust than others (Figure~\ref{fig:clustbench_k_misspecification}). 

Similarly, we ported the SVG benchmark \cite{bioRxiv2023spatially_variable_genes} using the highly modular \texttt{OpenProblems} \cite{openproblems} implementation, integrating both the benchmark plan and most of the modules (methods and metrics) within a single code repository (simplified plan shown in Figure \ref{fig:example_benchmarks}F). We downloaded the semi-simulated data and truth annotations from \texttt{OpenProblems} as well as their performance results, to characterize the consistency of results across benchmarks. We found a high reproducibility, with some methods (e.g., SpaGCN, SpatialDE2) showing a modest degree of variation (Fig.~\ref{fig:example_benchmarks}G) despite top-level package versions being the same.

In summary, \texttt{Omnibenchmark} is currently the only benchmarking system that seamlessly transitions between solo and collaborative benchmarking, is actively maintained, tested via unit and integration tests and is open to contributions (Extended Data~\ref{sec:architecture}). We are currently improving the benchmark formalization with a dedicated ontology building on \texttt{PROV-O} \cite{lebo2013prov}; extending the validations (e.g., input and output files formats and shapes, benchmark plans, consistency across runs); increasing support for \texttt{arm64} architectures; providing hooks to snapshot full benchmarks and automate submission to public repositories (e.g., Zenodo); and deploying automated benchmarking reports beyond \texttt{bettr}.

\backmatter

\bmhead{Data availability}

An archive of benchmark artifacts for \texttt{omni-clustering-benchmarks} and \texttt{omni-OpenProblems-svg} are available at Zenodo (DOI: 10.5281/zenodo.18598405).

\bmhead{Code availability}

We provide tutorials, how-to guides and a reference at \url{https://omnibenchmark.org}. \texttt{Omnibenchmark} is developed at \url{https://github.com/omnibenchmark/omnibenchmark}. Benchmark plans for the \texttt{omni-clustering-benchmarks} and \texttt{omni-OpenProblems-svg} are available at \url{https://github.com/omnibenchmark/clustering_example} (branch \texttt{update-to-0.4-full}) and \url{https://github.com/A-meara/spatially_variable_genes}, respectively; code to reproduce the analysis and figures shown here is stored at \url{https://github.com/omnibenchmark/omnibenchmark_paper_code}.

\bmhead{Acknowledgments} 

We thank Maruša Koderman for the excellent visual design of Figure 1. We thank the FOSS community for providing the building blocks of \texttt{Omnibenchmark} (\texttt{EasyBuild}, \texttt{Git}, \texttt{Snakemake}, \texttt{Conda}, \texttt{Apptainer}, etc). We thank the Renku team at the Swiss Data Science Center (SDSC) and Robinsonlab members for their constructive feedback. We also thank colleagues that participated in earlier variants of the project or in hackathons and in particular, the participants of a Benchmarking PhD course at the University of Copenhagen in November 2025 for their feedback. 

\bmhead{Author contributions}

Conceptualization: IM, AL, BC, AS, MDR. Methodology: all. Software: IM, AL, DI, BC, RG, MDR. Benchmarking: IM, AM. Original draft preparation: IM. Visualization: IM, AL, AM, MDR. Writing, reviewing and editing: all. Supervision: IM, MDR. Funding acquisition: MDR.

\bmhead{Funding}

MDR acknowledges funding from the Swiss National Science Foundation (grants 200021\_212940 and 310030\_204869) as well as support from swissuniversities P5 Phase B funding (project 23-36\_14). CS is supported by the Novartis Research Foundation.

\bmhead{Conflict of interest}

The authors declare no conflicts of interest.

\clearpage

\begin{appendices} %%%%%%%%%%%%%%%%%%%%%%%%%%%%%

%\include{supplementary_data_ob}
% this is sourced as an Extended Data at `article.tex`, hence a bare skeleton

% S-staring numbering
\renewcommand{\thefigure}{S\arabic{figure}}
\renewcommand{\thetable}{S\arabic{table}}

{\Large Extended Data for:  Omnibenchmark: transparent, reproducible, extensible and standardized orchestration of solo and collaborative benchmarks}

\vspace{1.5em}

\begin{center} 

\noindent Izaskun Mallona$^{1,2}$, Almut Luetge$^{3}$, Ben Carrillo$^{1,2}$, Daniel Incicau$^{1}$,  
Reto Gerber$^{1,2}$, Aidan Meara$^{1}$, Anthony Sonrel$^{1,2}$, Charlotte Soneson$^{2,4}$,  
Mark D. Robinson$^{1,2}$

\vspace{1.5em}

\noindent $^{1}$ Department of Molecular Life Sciences, University of Zurich, Zurich, Switzerland \\
$^{2}$ SIB Swiss Institute of Bioinformatics, Zurich, Switzerland \\
$^{3}$ Swiss Data Science Center, Zurich, Switzerland \\
$^{4}$ Friedrich Miescher Institute for Biomedical Research, Basel, Switzerland
\end{center}

\tableofcontents

%\Extended Data

\pagebreak

\section{Supplementary Note: \texttt{Omnibenchmark} software}
\label{sec:sup_methods}

\subsection{Architecture and implementation}
\label{sec:architecture}

\subsubsection{Design}

The codebase (available at \url{https://github.com/omnibenchmark/omnibenchmark}) is structured as a modular Python package, with the core API providing access to benchmark plan handling, workflow orchestration, and reporting. An interactive description of Omnibenchmark's architecture is given at \url{https://docs.omnibenchmark.org/latest/architecture/}. The API includes functions to define and track datasets, describe benchmark tasks, and launch workflows through \texttt{Snakemake} \cite{koster2012snakemake}. Configuration is managed via \texttt{YAML} files (i.e., benchmark plans) that represent the (paths to) files to be generated by the workflow, and software backends (e.g., modules, containers, or conda environments). Artifact provenance tracing is explicit and encoded by the file directory structure, with nested folders depicting lineage dependencies.

The CLI, invoked through the \texttt{ob} command, is the main interface for users. It provides subcommands for initializing and validating benchmark plans, templating benchmarking modules, running benchmarks, collecting results, and generating summary reports (the full list of current commands and subcommands can be found in Table~\ref{tab:commands}). Each subcommand is documented with comprehensive \texttt{--help} output. The CLI integrates tightly with Snakemake and its plugin ecosystem, allowing users to execute benchmarks in controlled software environments. Hence, benchmarking runs can be reproduced consistently across different systems, from local machines to high‑performance computing clusters, by decoding the benchmark plan and running its components in reproducibly-built software environments.

The reporting functionality of Omnibenchmark allows aggregating performance metrics into standardized reports, which include performance metrics, links to code and data, as well as authorship attribution. The CLI also provides a facility (\texttt{ob dashboard --format bettr}) to create inputs compatible with the interactive R/Shiny-based metric exploration tool named \texttt{bettr}  (accessible at \url{https://bettr-server.shinyapps.io/bettr-server/}).

\subsubsection{Test suite}

The codebase includes a continuous integration pipeline that orchestrates unit and integration tests on pull request. The test suite covers the CLI commands, benchmark plan handling and parsing, and workflow orchestration. Dedicated test environments (e.g., \texttt{omni-environment.yml} or \texttt{pixi.toml}) are provided to ensure software environment isolation and robustness across Python versions and other dependencies. 

End-to-end tests using example benchmarks with known layouts and expected outputs are also included. These examples include malformed plans and failing or misspecified modules, hence testing validation, \texttt{YAML} benchmark plans parsing, module execution order and exit status, and compliance in output files location and content.

\subsubsection{Contributing}

Bug reports, feature requests and other issues are welcome at \url{https://github.com/omnibenchmark/omnibenchmark/}. Code or architectural contributions are also welcomed; we describe the procedure to set up a reproducible development and testing environment within the \texttt{CONTRIBUTING.md} file. 

\subsubsection{Installation}

\texttt{Omnibenchmark} is implemented in Python and distributed under the free software Apache 2 license. The package is available through the Python Package Index (PyPI) and can be installed using the standard Python package manager \texttt{pip} with \texttt{pip install omnibenchmark}. The codebase includes a \texttt{uv.lock} to ensure deterministic dependency resolution using \texttt{uv}.

The \texttt{Omnibenchmark} package is designed to interoperate with \texttt{Snakemake} (workflow orchestration), \texttt{Lmod} (environmental modules system), \texttt{Apptainer} (container runtime), and \texttt{Conda} (package and environment manager). These tools are not required for installing the Python package itself, but they are essential for executing benchmarks in a reproducible manner; particularly, \texttt{Conda} is recommended to aid in the \texttt{Snakemake} installation. Hence, installing \texttt{Omnibenchmark} inside a dedicated \texttt{Conda} environment (preferably with Miniforge) is recommended. Instructions to make use of reproducible software distributions, such as \texttt{EESSI} \cite{droge2023eessi} and other repositories based on CernVM-FS \cite{blomer2013cernvm,blomer2017new} are provided.

\subsubsection{Computational performance profiler}

\texttt{Omnibenchmark} makes use of the performance metrics recorded by \texttt{Snakemake}, including wall-clock time, CPU time, peak memory consumption, cumulative memory usage, and I/O statistics. 

\subsubsection{Documentation}

The codebase includes technical documentation as well as user-facing documentation, with tutorials, how-tos, and a CLI reference. Documentation is versioned, bundled to the CLI releases, built via continuous integration on release, and deployed to \url{https://docs.omnibenchmark.org/}.

\subsection{Building components, running and sharing benchmarks}
\label{sec:operations}

\subsubsection{Structure of the benchmark \texttt{YAML} plan}
\label{sec:yaml}

\texttt{Omnibenchmark} offers a formal specification of benchmarking components that is serialized as a \texttt{YAML} file containing a header (describing general aspects of the benchmark such as metadata and storage) and a body (specifying benchmarking stages, and \emph{modules} and their dependencies).

The header of the \texttt{YAML} file serves as the preamble and declares the benchmark identifier, description, version, and maintainer contact. It also specifies the storage API (e.g., S3-compatible for sharing results) and the software environments required: each environment is described by a conda specification, a container image or environment module (\emph{envmodule}; this naming convention aims to avoid ambiguity with \emph{modules}, which refer to the repositories implementing benchmarking steps) name, and a short description of these. This ensures that the benchmark can be executed reproducibly across computing infrastructures. The header also declares the \texttt{YAML} schema version, to keep track of cross-compatibility with different versions of \texttt{Omnibenchmark}.

The body of the YAML is organized into \emph{stages} (coarse grained) and \emph{modules} (fine grained). Each stage groups modules that share input and output data shapes and formats. For example, a \emph{data-simulation} stage may define one or more dataset modules, each pointing to a \texttt{Git} repository and a commit that generates a dataset shaped as a CSV with features as rows and samples as columns. Then, a \emph{methods} stage may collect one or more modules consuming the outputs of the data stage and producing standardized result files (e.g.,  sample embeddings in a CSV format where samples correspond to rows and columns to coordinates). Finally, a \emph{metrics} stage would contain one or more modules ingesting the outputs from the \emph{method} stage modules and producing performance evaluation results (e.g., JSON files). While dataset-method-metric stages is the canonical design, \texttt{Omnibenchmark} plans can accommodate more flexible topologies. Each module is defined by a unique identifier, a software environment, a repository URL, a commit hash, and given the stage, explicit input/output specifications. Wildcards such as \texttt{\{dataset\}} (representing the identifier of a module from the first stage) are used in output paths to automatically organize workflow outputs into a hierarchical directory structure, preserving provenance. To collect results stored within different (parallel) directories (i.e., generated from different dataset lineages), the \texttt{YAML} can be used to define one or more \textit{metric collectors}, which aggregate outputs from multiple modules into a single report. This is particularly useful for summarizing results across datasets and methods. Metric collectors are defined as a module, with their repository and environment; their main inputs are the outputs from the metrics stage modules.

Benchmark plans can be templated using \texttt{ob create benchmark}. Once generated, these can be validated with the \texttt{ob validate plan} command, ensuring consistency before execution.

A full Benchmark Definition Reference can be found in the documentation at: \url{https://docs.omnibenchmark.org/latest/spec-reference/}, which shows all available options, their types, and whether they are required.

\subsubsection{Module structure}
\label{sec:modules}

In practice, modules need to follow the stage validation/structure contract as defined by the benchmark plan. That is, given all modules from a stage must consume the same inputs and produce defined outputs, the code from a benchmarking module must be able to implement their relevant algorithm that fulfils such input/output contracts. Similarly, modules must be able to run within the software environment assigned to them within the benchmark plan. Module contributors thus must organize and test both components.

Any Git‑accessible repository can serve as a module and should provide a license, a \texttt{CITATION.cff} file listing its author(s), configuration file and an entrypoint script compatible with the standardized arguments \texttt{Omnibenchmark} injects at runtime: namely, \texttt{--output\_dir}, \texttt{--name},  any declared inputs, and parameters. Compliance with these requirements can be checked using \texttt{ob validate module}. Outputs must be written exactly as specified in the benchmark plan, so that downstream modules can locate them.

Module contributors decide on the module granularity, i.e., whether a single \texttt{Git} repository implements a single method (dataset, metric, stage, etc.) or whether it dispatches multiple. Modules compatible with a given benchmark plan can be scaffolded using \texttt{ob create module}. A browser-based benchmark plan editor (and validator) is available from: \url{https://omnibenchmark.github.io/obeditor/}, with a tab for the user to insert the benchmark plan; other tabs will show a graphical representation of the benchmark topology; tabular summary of stage inputs and outputs; an explicit \texttt{Snakefile}, and options to filter and share results from a running benchmark.  

\subsubsection{Refining execution and data sharing using Snakemake plugins}

By default, running a benchmark with \texttt{ob run plan.yaml} triggers the execution of a Snakemake workflow locally (\texttt{--dry} does a dry run) . Adding the argument \texttt{--executor slurm} submits jobs to a Slurm-managed queuing system instead. Other remote executors can be added by leveraging the Snakemake executor plugin ecosystem (https://snakemake.github.io/snakemake-plugin-catalog/).

Storing results in remote S3-compatible storage involves  enabling  valid storage remotes (i.e., a S3-compatible bucket), that can be managed with \texttt{ob remote} (Table~\ref{tab:commands}). Further remote storage options (Dropbox, Azure, FTP, etc.) can be opted in by embracing the Snakemake storage plugin ecosystem (https://snakemake.github.io/snakemake-plugin-catalog/).

To facilitate extensibility beyond storage and execution, further Snakemake-specific arguments can be appended to \texttt{ob} CLI calls by appending a \texttt{--} followed by the desired (Snakemake-specific) extra arguments.

\subsubsection{Reusing a benchmark or its components}

Downloading benchmarking outputs (or files) using the associated benchmark plan enables users not only to run the benchmark from scratch, but also to modify the plan (i.e., add, modify or delete components) and in principle, only run the newly-introduced steps to bring the benchmarking outputs to a consistent state with the updated plan. However, this may only be a mode of operation in collaborating on or contributing to a benchmark, since the results would need to be run on homogeneous hardware for final results.

We are aware that, beyond benchmarking, modules and software environments have a value to end users. Omnibenchmark modules include a mandatory CFF file with citation details, an explicit license and an \texttt{ob cite} command to facilitate the collection of intellectual attribution when reusing or modifying benchmark components.

\subsubsection{Archiving (snapshotting) a benchmark}

\texttt{ob archive} has various options to produce reproducible and portable snapshots of the current state of a benchmark and/or its outputs. Users can select the contents to be archived, including the benchmark plan, code repositories, software environments, and/or result files. Archives can be compressed with different methods and compression levels (uncompressed, deflated, bzip2, lzma, or gzip). Archives can be created from local directories or, when enabled, by retrieving results from remote storage. A dry-run mode displays the files to be archived without generating the snapshot.

\subsubsection{Attributing and citing a benchmark}

Similarly, the mandatory CFF file in each benchmarking module ensures that citation files (e.g., BibTeX) can be generated for any benchmark using \texttt{ob cite}. Hence, the attribution of contributors of all modules, plus the authors listed in the benchmark plan, can be aggregated.

\subsection{Solo vs collaborative benchmarking: distribution and centralization strategies}
\label{sec:centralization}

In a \emph{solo} scenario, a single researcher can design and run a benchmark entirely locally. Here, the user drafts a benchmark plan and implements the modules for datasets, methods, and metrics, and executes the workflow on a personal machine using reproducible environments. All modules, data, and results can remain private, with Git repositories and container recipes tracked locally. This mode is suited for exploratory or confidential benchmarking, where reproducibility is ensured, but sharing or seeking contributions is not required.

At the other end of the spectrum, \texttt{Omnibenchmark} enables open, community‑based benchmarking. In this model, benchmark plans and modules are hosted on public repositories (e.g., GitHub), and continuous execution is orchestrated on shared infrastructure (e.g., EUDAT). Contributions of new datasets, methods, or metrics are made via pull requests, and results are automatically regenerated and disseminated. This approach leverages the transparency from FAIR principles; the long‑term sustainability (in terms of refining and releasing new versions) depend on the governance and community engagement level.

\texttt{Omnibenchmark} also accommodates communities that wish to restrict some freedoms. For example, communities may host private repositories on institutional GitLab servers, run benchmarks on HPC clusters, and share results selectively via authenticated object storage (e.g., Amazon S3). This mode is particularly useful to protect sensitive data, intellectual property, or unpublished methods, while still embracing extensibility and reproducibility. 

In all cases, the benchmark plan provides the common grammar to specify the benchmark components and layout, ensuring that benchmarks can be inspected, extended, and run, assuming no credentials nor access control are needed to have access to resources.

\subsubsection{Continuous integration}

To automate the validation checks and benchmarking (re)runs upon contribution, \texttt{Omnibenchmark} provides a CI/CD action (\url{https://github.com/marketplace/actions/run-omnibenchmark}) to trigger benchmarks on demand or upon changes on benchmark plans. CI/CD runners can be registered in any computing infrastructure, facilitating execution on self-managed servers.

\subsubsection{Consequences of a federated contribution with centralized coordination}

The \texttt{Omnibenchmark} design is largely federated and resembles distributed version control logic. Individual groups define and maintain their own modules in independent repositories with full autonomy over implementation and updates. These modules remain usable in isolation, but once registered in a benchmark plan, they are executed within a shared infrastructure. This semi-centralized arrangement ensures that contributions retain their provenance while still being harmonized into a reproducible workflow, while also being expected to evolve separately.

This federated contribution is also similar to open‑source package registries such as R/Bioconductor: developers work in a distributed manner producing software packages that, once ready and after community inspection, are deposited in a central registry. In \texttt{Omnibenchmark}, the benchmark plan and execution and data sharing infrastructure play the role of that registry: they do not override ownership of the modules, but they build an ecosystem out of them. This `develop anywhere, converge on execution' model enables high flexibility, transparency, and component reuse, giving value to the benchmarking components beyond benchmarking results, including: module code (e.g., human-curated and functional algorithm implementations), and software environments (to ensure the benchmarking modules are runnable and that installation of their dependencies is feasible and reproducible).

\clearpage

\section{Supplementary Note: Extending an existing Omnibenchmark}
\label{sec:extend}

To show that an \texttt{Omnibenchmark} can be easily extended while reusing pre-computed and compatible results without recalculating them, we started from the completed \texttt{omni-clustering-benchmarks} result set (Zenodo DOI: 10.5281/zenodo.18598405; described in Supplementary Note~\ref{supp:clustbench}). We added a new clustering module, a random-label partitioner that was not part of the original benchmark (to be interpreted as a baseline method), updated the benchmark plan, and re-ran it. Only the new method and the downstream metric steps were recomputed.

This mechanism is also the basic currency in multi-team collaborative benchmarks: they are extended by adding a new dataset, method, or metric as an independent (or updated) module, and a request is sent to update the benchmark plan to include them. 

\subsection{Steps}

First, we created a new module compatible with the \texttt{clustering} stage. Its input/output contract (consuming \texttt{data.matrix} and \texttt{data.true\_labels}, producing \texttt{clustering.predicted\_ks\_range}) is fixed by the plan:

\begin{lstlisting}[language=bash,numbers=none]
ob create module
\end{lstlisting}

The module wraps the new code behind a command-line interface (e.g., via \texttt{argparse}) that reads \texttt{--data.matrix} and \texttt{--data.true\_labels}. Notice these \texttt{data.matrix} and \texttt{true\_labels} are encoded verbatim as in the benchmark plan. For each tested number of clusters (the true $k$ plus or minus two), it assigns every point a uniformly random label, and writes one labels file per dataset:

\begin{lstlisting}[language=python]
#!/usr/bin/env python
"""Random-label baseline partitioner for the clustbench clustering stage."""
import argparse, os
import numpy as np

def generate_k_range(k):
    return dict(enumerate(max(x, 2) for x in (k - 2, k - 1, k, k + 1, k + 2)))

def main():
    p = argparse.ArgumentParser(description='clustbench random partitioner')
    p.add_argument('--data.matrix', required=True)
    p.add_argument('--data.true_labels', required=True)
    p.add_argument('--output_dir', required=True)
    p.add_argument('--name', default='clustbench')
    p.add_argument('--seed', type=int, default=123)
    args = p.parse_args()

    truth = np.loadtxt(getattr(args, 'data.true_labels'), ndmin=1)
    X = np.loadtxt(getattr(args, 'data.matrix'), ndmin=2)
    Ks = generate_k_range(int(max(truth)))

    rng = np.random.default_rng(args.seed)
    preds = np.array([rng.integers(1, K + 1, size=X.shape[0])
                      for K in Ks.values()]).T

    header = np.array(['k=%s' % s for s in Ks.values()]).reshape(1, 5)
    out = np.append(header, preds.astype(str), axis=0)
    np.savetxt(os.path.join(args.output_dir, f'{args.name}_ks_range.labels.gz'),
               out, fmt='%s', delimiter=',')

if __name__ == '__main__':
    main()
\end{lstlisting}

Second, because it needs only \texttt{numpy}, we reused the existing \texttt{clustbench} software environment (that the \texttt{sklearn} module already uses), instead of building a new one.

Before wiring it into the plan, we tested the module locally on a single dataset, pointing at data produced by an earlier run:

\begin{lstlisting}[language=bash,numbers=none]
python run_random.py \
  --data.matrix /home/user/clustering_example/out/data/clustbench/atom/atom.data.gz \
  --data.true_labels /home/user/clustering_example/out/data/clustbench/atom/atom.labels0.gz \
  --output_dir /tmp/random_test --name atom
\end{lstlisting}

This writes \texttt{/tmp/random\_test/atom\_ks\_range.labels.gz} in the format expected by the \texttt{metrics} stage.

Then, we updated the benchmark plan by adding one stanza to the \texttt{clustering} stage of \texttt{Clustering\_conda.yml}, alongside the existing \texttt{fastcluster}, \texttt{sklearn}, \texttt{agglomerative}, \texttt{genieclust} and \texttt{fcps} modules:

\begin{lstlisting}[language=bash,numbers=none]
      - id: random
        name: "random partitioner"
        software_environment: "clustbench"
        repository:
          url: /home/user/clustbench_random
          commit: main
        parameters:
          - seed: 123
\end{lstlisting}

During development, we often set \texttt{commit: main} so that the module tracks the branch tip. The URL can also point to a local path, hence not having to push the code to a remote (see comment below about running in-development benchmarks). This avoids committing and pinning a new hash on each iteration, which speeds up testing. On success, the merge request would pin \texttt{commit:} to a specific hash and point \texttt{URL} to a Git-compatible publicly available remote.

Next, we validated the module and the updated plan with \texttt{ob}:

\begin{lstlisting}[language=bash,numbers=none]
ob validate module
ob validate plan Clustering_conda.yml
\end{lstlisting}

Then, we did a dry run to confirm that Snakemake built a workflow scheduling only the new \texttt{random} module, plus the downstream metric steps.

\begin{lstlisting}[language=bash,numbers=none]
ob run Clustering_conda.yml --dry --dirty
\end{lstlisting}

\textit{Note: Running in-development benchmarks}. The benchmark, as specified above, can only be (dry) run with \texttt{--dirty} when local directories are used and with \texttt{--unpinned} when branch names are specified. Both flags emphasize that these runs may not be reproducible.

Finally, we ran the plan, which computed only the scheduled jobs:

\begin{lstlisting}[language=bash,numbers=none]
ob run --cores 50 --continue-on-error Clustering_conda.yml
\end{lstlisting}

\subsection{Incremental compute}

\texttt{Omnibenchmark} encodes artifact provenance in the output directory structure, with nested folders following the dataset-method-metric lineage. Adding a method extends the workflow's directed acyclic graph (creating \texttt{random} paths for every dataset, and the matching \texttt{partition\_metrics} outputs downstream).

The output files for the \texttt{random} module and its downstream metrics do not yet exist, so the created \texttt{Snakemake} workflow triggered by \texttt{ob run} schedules them. Other files present on the machine, e.g., results from the other clustering methods, or the datasets, are unchanged, so \texttt{Snakemake} marks them up to date and skips them. This output reusage mechanism applies regardless of whether results are local or in remote storage, either S3-compatible or through other \texttt{Snakemake} storage plugins.

\subsection{Multi-team collaboration}

The same procedure of modifying modules and updating the plan, performed by multiple contributors according to a commonly-agreed plan, can be achieved within a collaborative benchmark. Because artifacts from unaltered modules or stages can be reused across teams, contributors can add components independently without recomputing each other's work. 

Shared modules, in particular datasets, can be reused across several benchmarks, perhaps even seeded from one scaffold. We apply this model in ongoing multi-team single-cell benchmarks, including \texttt{omni-scrna} and \texttt{omni-scbench}, which are listed among other community benchmarks at \url{https://docs.omnibenchmark.org/latest/examples/#community-benchmarks} (the list is actively updated). \texttt{omni-scrna} is a single-cell scaffolding benchmark aimed for reusability across multiple (single-cell RNA-seq related) tasks. It was populated by roughly twelve computational biologists in task-specific teams during a community hackathon. To add governance and automation, we made use of multiple layers of CI/CD (e.g., GitHub Actions): i) to \texttt{ob validate} newly-committed modules or benchmark plans; ii) to implement quick ``smoke tests'' that run new modules on a constructed small dataset; iii) to run the full benchmark (using a local runner) such that outputs can be shared with the team. These automations are built into the repository-level GitHub Actions, making use of the \texttt{run\_omnibenchmark} routine provided at the GitHub Marketplace (\url{https://github.com/marketplace/actions/run-omnibenchmark}), which installs \texttt{Omnibenchmark} and runs a plan under a chosen software backend (conda or Apptainer). 

%Execution can be run either on a GitHub runner or in custom hardware, provided they install and register a GitHub runner executor. 

At time of writing (August 2026), \texttt{omni-scrna} is actively growing. We organize contributions into contributor subgroups using GitHub Projects, with each new dataset, method, or metric entering via pull requests to the plan repository, tested by multiple means (GitHub Actions plus often local tests). To keep modules interchangeable within a stage, we define schemas for the stage-wise command-line arguments, so that every module in a stage reads and writes the same inputs and outputs. 

%The schemas are checked with \texttt{ob validate}. 

%As a safeguard, each pull (merge) request triggers continuous integration that validates the plan and the changed modules. It also dry runs the whole benchmark plan, as well as a small-sized benchmark variant with small datasets, so a contribution is tested before it is merged. 

\clearpage

\section{Supplementary Note: \texttt{omni-clustering-benchmarks} Omnibenchmark}
\label{supp:clustbench}

Following the original \texttt{clustering-benchmarks} \cite{gagolewski2022framework}, we benchmarked 62 datasets with a known ground truth and 48 methods using 3 partition metrics. If present, we included ``noisy points'' during the clustering process, but ignored them to calculate the performance metrics.

We grouped datasets and methods according to their generator and/or software environment, hence writing modules able to dispatch multiple methods. In particular, the \texttt{Conda}-backed benchmark plan exposes parameters of some modules to dispatch running of certain methods/datasets. Beyond \texttt{Conda}, we also designed \texttt{EasyBuild}-built (extending EESSI \cite{droge2023eessi} 2025.06) and \texttt{Apptainer} execution environments to evaluate the impact of the software backend in benchmarking results, both in terms of algorithmic (e.g., clustering metrics) and computational (e.g., speed, memory fingerprint) performance.

In the original publication, Gagolewski \cite{gagolewski2022framework} reported the aim of \texttt{clustering-benchmarks} to provide a systematic and diverse catalogue of datasets of different sizes, dimensions and complexity, as well as to use a curated set of metrics, while also highlighting that there are multiple equally-valid ways to cluster a given dataset. The ported \texttt{omni-clustering-benchmarks} Omnibenchmark does not aim to duplicate these efforts to re-evaluate method performance, but, rather, focuses on exploring the impact of computational performance according to the software installation strategy.

We ran benchmarks on a Linux machine with x86\_64 architecture, zen2 AMD EPYC 7742 64-Core processor.

\section{Supplementary Note: Investigation of Adjusted Rand Index discrepancy determined by compiler flags}
\label{sec:wut-z1-ward}

Across seeds and repetitions, one path in the \texttt{omni-clustering-benchmark} reports a different Adjusted
Rand Index under the EasyBuild backend than under conda and Apptainer: the
\texttt{fastcluster} module with Ward linkage on the \texttt{wut/z1} dataset
(Figure~2C). It is the only such discrepancy in the benchmark.

\begin{table}[h]
\centering
\begin{tabular}{lccc}
\toprule
Method & conda & Apptainer & EasyBuild \\
\midrule
\texttt{fastcluster/linkage-ward}   & 0.2737328 & 0.2737328 & 0.1443372 \\
\texttt{agglomerative/linkage-ward} & 0.1835665 & 0.1835665 & 0.1835665 \\
\bottomrule
\end{tabular}
\caption{Adjusted Rand Index at $k=3$ on \texttt{wut/z1}, identical across all
three seeds and both repetitions.}
\label{tab:wut-z1-ari}
\end{table}

All three environments pin the python package \texttt{fastcluster} 1.2.6; the difference lies in the build rather than the source code. conda and Apptainer install it from PyPI as a manylinux wheel, compiled for the x86-64 baseline because it must run on unknown hardware; EasyBuild compiles on site and applies \texttt{-march=native} by default. Where fused multiply-add instructions are available, GCC's default
\texttt{-ffp-contract=fast} evaluates $a b + c$ with one rounding instead of two. This is the shape of the quantity that Ward minimizes in \texttt{fastcluster} (multiply-add accumulations), so the results shift by roughly one unit in the last place between the two builds.

This numerical shift propagates to a divergent result on \verb|wut/z1| because its geometry yields exactly equal pairwise distances. FMA optimizations perturb these values at the last bit, altering arbitrary tie-breaking in Ward's merges. We consider this case a degenerate dataset rather than a backend reproducibility failure: better version pinning cannot fix undefined tie-breaking behavior.

A self-contained reproduction, which builds \texttt{fastcluster} 1.2.6 from the same sdist under eight flag settings and reports the FMA instruction count and resulting Adjusted Rand Index for each, is available at
\url{https://gist.github.com/btraven00/7c27ff9320adf00f35f516f158c50f54}.

\section{Supplementary Note: \texttt{omni-OpenProblems-svg} Omnibenchmark}

To highlight the adaptability of the Omnibenchmark framework, we ported the Spatially Variable Genes (SVG) task from Open Problems (see also \cite{bioRxiv2023spatially_variable_genes}), which aims to evaluate the selection of genes whose expression vary across spatial locations within spatial transcriptomic data (e.g., slices of tissues or organ with underlying spatial structure). We reused the input data (e.g., including simulated ground truths) and adapted their software environments and scripts to run methods and metrics as Omnibenchmark modules. We then compared the performance metrics from the ported Omnibenchmark to the original results.

In terms of evaluation, Li et al \cite{bioRxiv2023spatially_variable_genes} built synthetic data with controlled and known amounts of spatial or non-spatial signal, and quantified performance via Kendall's $\tau$ correlation (a measure of agreement) of true and predicted spatial gene expression variability. Hence, a task-specific performance is summarized with a single score. We also profiled computational performance (data not shown).

Given the high modularity of Open Problems benchmarks and its \texttt{Viash} components \cite{cannoodt2024viash}, we largely maintained the original SVG dataset, method and metric layout, except for adding boilerplate code to make them callable via command-line via argument parsing. Similarly, we ported the \texttt{Viash} stanzas addressing software environments to Apptainer recipes to make them compatible with \texttt{Omnibenchmark}. Software was ported with minimal changes, keeping package versions unpinned if they were in the \texttt{Viash} specification.

The benchmark plan contained 50 datasets, two control methods, 14 methods and a single (Kendall's $\tau$) metric and is available at Zenodo (see `Data availability'), including the required accessors (e.g., not benchmarked steps) to retrieve simulated data and metric results from Open Problems.

We ran the benchmark on a Linux machine with x86\_64 architecture, zen2 AMD EPYC 7742 64-Core processor.

\clearpage

\section{Supplementary Tables}

\begin{table}[!ht]
  \centering
  \caption{Non-exhaustive overview of topic-specific benchmarking frameworks.}
  \label{tab:topic-specific}
  \footnotesize
  \begin{tabularx}{\linewidth}{
    >{\hsize=.65\hsize\raggedright\arraybackslash}X
    >{\hsize=1.15\hsize\raggedright\arraybackslash}X
    >{\hsize=.95\hsize\raggedright\arraybackslash}X
    >{\hsize=1.1\hsize\raggedright\arraybackslash}X
    >{\hsize=.55\hsize\raggedright\arraybackslash}X
  }
    \toprule
    Framework & Target task or scope & Implementation & Method or dataset contribution & Continuous evaluation \\
    \midrule
    \rowcolor{gray!12} scIB \cite{scib}             & Single-cell data integration (batch correction)          & Python package and \texttt{Snakemake} & Wrap methods behind the \texttt{scib} API         & No  \\
    CellBench \cite{cellbench}   & Combinations of single-cell RNA-seq analysis methods     & R/Bioconductor                        & Functionally compose pipeline steps               & No  \\
    \rowcolor{gray!12} DANCE \cite{dance}           & Single-cell and spatial deep-learning tasks              & Python (PyTorch)                      & Implement models in the DANCE API                 & No  \\
    dynverse \cite{dynverse}     & Trajectory inference                                     & R packages and containers             & Containerized method wrappers                      & No  \\
    \rowcolor{gray!12} LEMMI \cite{seppey2020lemmi} & Metagenomics classifiers                                 & Web platform and containers           & Submit a containerized tool                        & Yes \\
    BenchHub \cite{benchhub}     & Results-linking ecosystem (computational biology)        & R6 ``Trio'' database                  & Submit datasets, metrics and results               & Yes \\
    \bottomrule
  \end{tabularx}
\end{table}

\begin{table}[ht]
  %\centering
  \caption{Comparison of general-purpose community benchmarking platforms. Frameworks built for specific tasks are listed in Table~\ref{tab:topic-specific}.}
  \footnotesize
  \begin{tabularx}{\linewidth}{
    >{\hsize=.4\hsize\raggedright\arraybackslash}X  % Macroproperty
    >{\hsize=.6\hsize\raggedright\arraybackslash}X  % Property
    >{\hsize=1.25\hsize\raggedright\arraybackslash}X  % Omnibenchmark
    >{\hsize=1.1\hsize\raggedright\arraybackslash}X  % Openproblems.bio
    >{\hsize=1.0\hsize\raggedright\arraybackslash}X  % OpenEBench
  }
    \toprule
                      & Property                        & Omnibenchmark                                                                              & Open Problems \cite{openproblems}                                                                       & OpenEBench \cite{openebench}                                                                \\
    \midrule
    \rowcolor{gray!12}\multirow{5}{=}{Technical features}
                                   & Workflow support                & Abstracted (self-generates a Snakemake workflow)                                          & \texttt{Viash} \cite{cannoodt2024viash}                                                                                   & Agnostic, e.g.\ Nextflow and others                                         \\
                                   & Compute location                & Not on premises, delegates execution to the user (local, cloud, HPC)                      & AWS, with automations                                                                   & On premises                                                                   \\
      \rowcolor{gray!12}           & Data and artifact storage       & Not on premises, delegates storage to the user (local, S3), git                           & On premises, all code and results available on GitHub                                   & On premises                                                                   \\
                                   & Software and containerization   & Supports Apptainer, Conda, Envmodules, or host environments via Snakemake integration     & Supports Docker and Apptainer via Nextflow integration                                  & Supports Docker and Apptainer; encourages ELIXIR's containerization best practices \\
      \rowcolor{gray!12}           & Platform or CLI                 & CLI                                                                                        & Platform                                                                                 & Platform                                                                      \\
    \midrule
    \multirow{3}{=}{Community and governance}
                                   & Maintainers                     & University of Zurich                                                                               & Helmholtz and Yale, with sponsors                                                      & ELIXIR (BSC)                                                                  \\
      \rowcolor{gray!12}           & Community engagement            & Delegates choosing the contribution model; open-source via git                             & Open-source contribution model with calls, hackathons, Discord                          & Open-source model coordinated with ELIXIR; events and hackathons              \\
                                   & Challenges and leaderboards     & Delegates organizing challenges; provides a dashboard to browse results (\texttt{bettr} \cite{bettr})           & Formalizes tasks and deploys continuous leaderboards                                    & Formalizes tasks and deploys continuous leaderboards                          \\
    \midrule
    \rowcolor{gray!12}\multirow{4}{=}{Formats and generalization}
                                   & File formats and standards      & Provides validation mechanisms and delegates decisions to users                           & Well-defined datasets and performance metrics; provides validation mechanisms           & Embraces FAIR formats and integrates with bio.tools                           \\
                                   & Generalizable metrics           & Installability, licensing, and performance profiling are built-in            & N/A                                                                                     & Buildability, installability, licensing, documentation are built-in            \\
      \rowcolor{gray!12}           & Benchmark scope (by design)     & Arbitrary, user-defined benchmarks; scientific-field agnostic                             & Arbitrary tasks (framework); deployed in single-cell genomics                           & Arbitrary tasks (framework); deployed across life sciences                    \\
                                   & Primary field of deployment     & Agnostic                                                                                  & Single-cell genomics                                                                    & Life sciences                                                                  \\
    \bottomrule
  \end{tabularx}
  \label{tab:comparison}
\end{table}

\begin{table}[ht]
%\centering
\caption{CLI commands with brief descripton.}
\begin{tabular}{l l}
\hline
{Command} & {Description} \\ \hline

\texttt{ob archive}& Archive a benchmark and its artifacts.\\
\texttt{ob cite}& Extract citation metadata from CITATION.cff...\\
\texttt{ob create benchmark}&  Create a new benchmark from a template.\\
\texttt{ob create module} & Create a new module from a template.\\ 
 \texttt{ob dashboard}&Generate a dashboard from benchmark results.\\
\texttt{ob describe snakemake}& Export computational graph to DOT format \\
\texttt{ob describe status}&Show the status of a benchmark\\
\texttt{ob describe topology}& Export benchmark topology to Mermaid diagram format \\
%\hline
\texttt{ob run}& Run a benchmark or a specific module.\\
\texttt{ob validate plan}& Validate benchmark YAML plan structure.\\
\texttt{ob validate module}& Validate module (metadata, try to run).\\

%\hline
\texttt{ob remote files}& Manage files in remote storage.\\
\texttt{ob remote policy}& Manage storage policies.\\
\texttt{ob remote version}& Manage benchmark versions.\\
\hline
\end{tabular}
\label{tab:commands}
\end{table}

\vspace{-2cm}

\begin{table}[ht]
    %\centering
            \caption{Software management strategies. }
\begin{tabular}{l c c c}
\hline
{Strategy} & {Linux} & {Windows} & {macOS} \\
\hline
Conda      & \checkmark & \checkmark & \checkmark \\
EasyBuild  & \checkmark & (partly)         & -- \\
Apptainer (Singularity) & \checkmark & --         & -- \\
\hline
\end{tabular}
%    \vspace{0.5cm}

\label{tab:software}
\end{table}

\clearpage %pagebreak

\section{Supplementary Figures}

\begin{figure}[hbt!]  
   \centering
      \includegraphics[width=\linewidth]{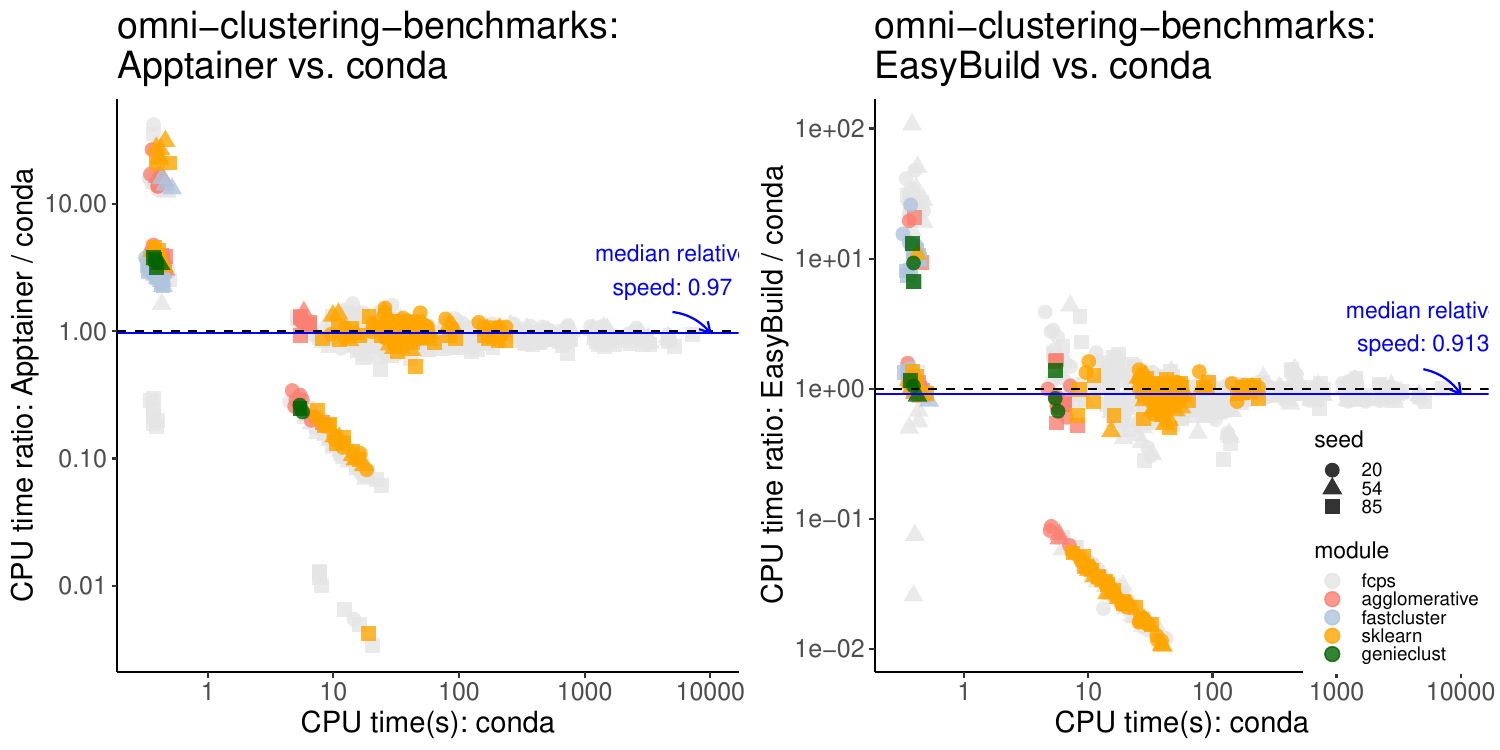} 
  \caption{Relative CPU time of \texttt{Apptainer} (left) and \texttt{EasyBuild} (right) software compared to \texttt{Conda} software versus \texttt{Conda} CPU time. All run times and ratios are presented here (filtering is applied in Figure \ref{fig:example_benchmarks})}
  \label{fig:cpu_time-supp}
\end{figure}

\begin{figure}[hbt!]  
   \centering
      \includegraphics[width=\linewidth]{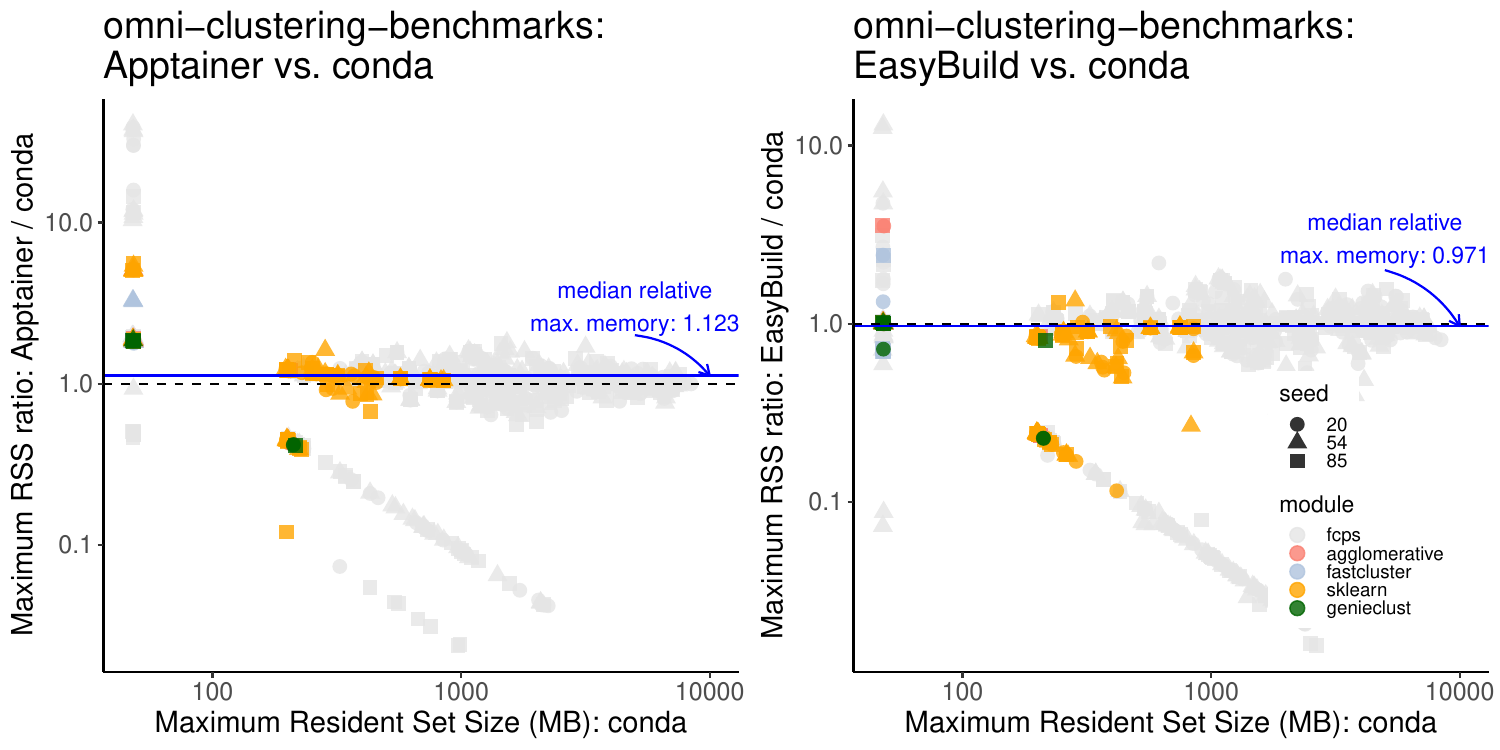} 
  \caption{Relative maximum Resident Set Size (RSS) of \texttt{Apptainer} (left) and \texttt{EasyBuild} (right) software compared to \texttt{Conda} software versus \texttt{Conda} maximum RSS (in megabytes). }
  \label{fig:max_rss-supp}
\end{figure}

\begin{figure}[hbt!]  
   \centering
      \includegraphics[width=\linewidth]{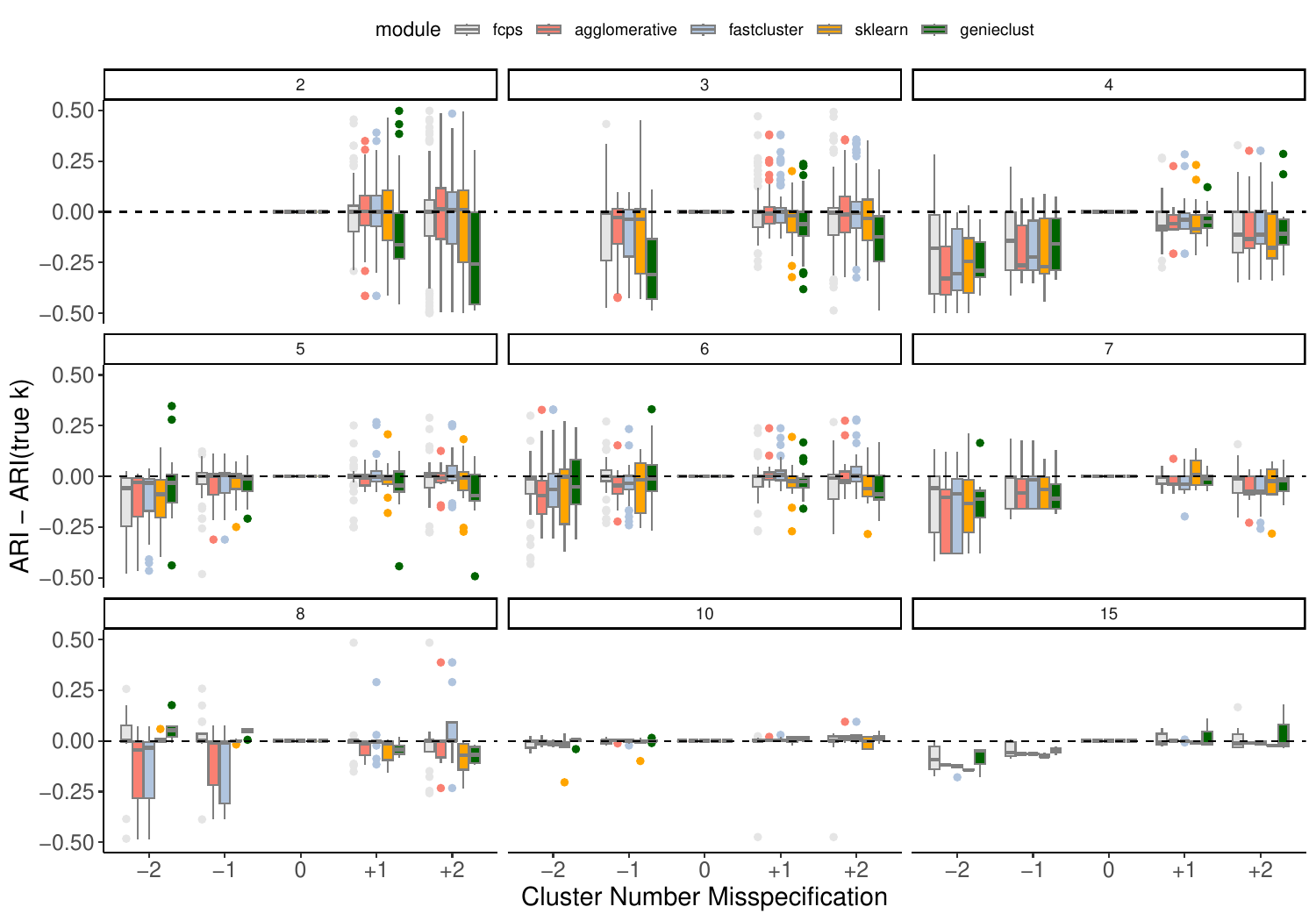} 
  \caption{The effort on algorithmic performance (here, change in Adjusted Rand Index relative to no cluster number misspecification) by misspecifying the number of clusters. Results are split by their \texttt{module} and faceted by the number of true clusters.}
  \label{fig:cluster-misspec-supp}
\end{figure}

\begin{figure}[hbt!]  
   \centering
      \includegraphics[width=\linewidth]{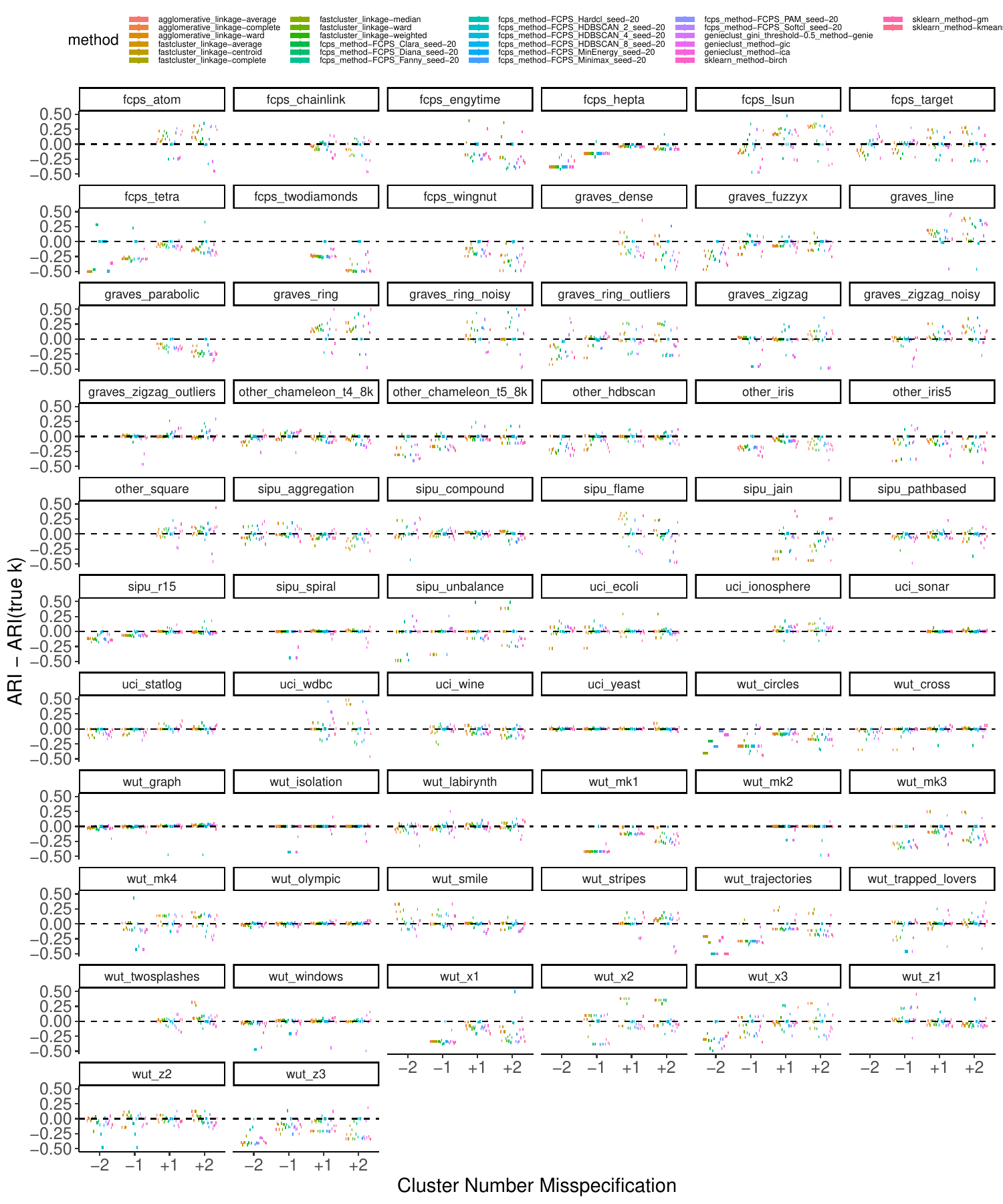} 
  \caption{The effort on algorithmic performance (here, change in Adjusted Rand Index relative to no cluster number misspecification) by misspecifying the number of clusters. Results are split by their \texttt{method} and faceted by dataset.}
  \label{fig:cluster-misspec-supp1}
\end{figure}

\end{appendices} %%%%%%%%%%%%%%%%%%%%%%%%%%%%%

\clearpage
\bibliography{bibliography}

\end{document}